\documentclass[journal=jacsat,manuscript=article]{achemso}

\usepackage[version=3]{mhchem} 
\usepackage{siunitx}
\usepackage{hyperref}
\usepackage{amsmath}
\usepackage{multirow}
\graphicspath{ {./Figures/} }

\newcommand*{\StudiedWires}{55,516~}

\author{Stephen A Church}
\affiliation[UoM]
{Department of Physics and Astronomy and The Photon Science Institute, The University of Manchester, M13 9PL, United Kingdom}
\email{stephen.church@manchester.ac.uk}
\author{Francesco Vitale}
\affiliation[Jena]
{Institute for Solid State Physics, Friedrich Schiller University Jena, Max-Wien-Platz 1, 07743 Jena, Germany}
\author{Aswani Gopakumar}
\author{Nikita Gagrani}
\affiliation[ANU]
{Department of Electronic Materials Engineering, Research School of Physics, The Australian National University, Canberra, ACT, 2601 Australia}
\author{Yunyan Zhang}
\affiliation[Zhejiang]
{College of Integrated Circuits, Zhejiang University, Hangzhou, Zhejiang 311200, China}
\author{Nian Jiang}
\affiliation[Cambridge]
{Department of Engineering, University of Cambridge, CB2 1PZ, United Kingdom}
\author{Hark Hoe Tan}
\author{Chennupati Jagadish}
\affiliation[ANU]
{Department of Electronic Materials Engineering, Research School of Physics, The Australian National University, Canberra, ACT, 2601 Australia}
\author{Huiyun Liu}
\affiliation[UCL]
{Department of Electronic and Electrical Engineering, University College London, WC1E 7JE, United Kingdom}
\author{Hannah Joyce}
\affiliation[Cambridge]
{Department of Engineering, University of Cambridge, CB2 1PZ, United Kingdom}
\author{Carsten Ronning}
\affiliation[Jena]
{Institute for Solid State Physics, Friedrich Schiller University Jena, Max-Wien-Platz 1, 07743 Jena, Germany}
\author{Patrick Parkinson}
\email{patrick.parkinson@manchester.ac.uk}
\affiliation[UoM]
{Department of Physics and Astronomy and The Photon Science Institute, The University of Manchester, M13 9PL, United Kingdom}

\title{Data-driven Discovery for Robust Optimization of Semiconductor Nanowire Lasers}


\begin{document}


\begin{abstract}
Active wavelength-scale optoelectronic components are widely used in photonic integrated circuitry, however coherent sources of light -- namely optical lasers -- remain the most challenging component to integrate. Semiconductor nanowire lasers represent a flexible class of light source where each nanowire is both gain material and cavity; however, strong coupling between these properties and the performance leads to inhomogeneity across the population. While this has been studied and optimized for individual material systems, no architecture-wide insight is available. Here, nine nanowire laser material systems are studied and compared using \StudiedWires nanowire lasers to provide statistically robust insight into performance. These results demonstrate that, while it may be important to optimise internal quantum efficiency for certain materials, cavity effects are always critical. Our study provides a roadmap to optimize the performance of nanowire lasers made from any material: this can be achieved by ensuring a narrow spread of lengths and end-facet reflectivities.
\end{abstract}


The miniaturisation of opto-electronic devices is an area of rapid development across multiple research fields~\cite{Park2015N,Zhang2019s,Xiong2021}. In its simplest form this involves the development of devices at the centi- and millimeter-scale, enabling previously bulky apparatus to become light-weight and portable, facilitating accurate field-based testing, with applications including quality control~\cite{Li2016N} and medical screening~\cite{Bansal2015N}. Further miniaturisation, to the micron-scale, has yielded significant advancements in commercial technology. This includes micro-LEDs~\cite{Wu2018N}, which form the foundation of ultra-efficient, high-resolution and wide colour-gamut display technologies~\cite{Zhou2015N}, enabling commercial virtual and augmented reality headsets~\cite{Xiong2021}.

Nanoscale coherent light sources are widely sought after as biological probes\cite{Fikouras2018, Wu2018a}, sensors, in quantum applications~\cite{Piranddola2020,Sutula2023} and most often as active components for photonic integrated circuitry\cite{Yang2023FromCircuits, Sun2015,Zhang2019s,Wang2020N} that integrate micron-scale light sources, waveguides and detectors at high densities to perform light-based computing processes~\cite{Shastri2021}. A wide range of architectures have been proposed as solutions, spanning off-chip coupling, flip-chip integration and hetero-epitaxial growth\cite{mayer2016a, Stettner2017, schuster2017}. However, repeatable, high-density and high-yield laser sources are rare. Of particular concern is the strong coupling between material quality, cavity design and performance, which mean that small variations in defect density, surface passivation, hetero-junction parameters and geometry lead to significant effects for performance\cite{Alanis2017,Liu2021QD,Wang2019QD}. 

Nanoscale lasers based on semiconductor nanowires (NWs) have long been proposed as coherent light sources that are well suited to integration\cite{Parkinson2021PhysicsLasers, eaton2016a, Couteau2015, Ning2010, Huang2004}; these provide diverse emission properties, such as wavelength, pulse duration and modal structure, arising from the wide choice of material system, while being united in their fundamental operating principle of the nanowire forming a monolithic cavity and gain material. These NW lasers (NWLs) typically have diameters of 100's nm and lengths up to 10's \(\mu\)m. High refractive indices result in strong optical confinement, and the narrow NW diameters often lead to a deterministic relationship between cavity, material and performance properties that necessitate optimization of each parameter~\cite{Church2022OpticalLasers}. The variety of material systems has led to a divergence in this optimization and development, and to-date high quality systems have been demonstrated in chalcogenides\cite{Bao2020, Li2013}, nitrides\cite{Li2017, Li2015c}, perovskites\cite{Dong2020,Schlaus2019}, metal oxides\cite{Sergent2020, Vanmaekelbergh2011} and III-Vs\cite{Church2023HolisticDesign, Tatebayashi2015, Saxena2013OpticallyLasers}. This range of systems provides significant advantages for selecting lasing wavelength, for high-speed operation\cite{Sidiropoulos2014} or providing specific non-linear properties\cite{Yi2022Self-frequency-conversionLasers}. However, it is difficult to robustly compare the performance of each material system due to differences in excitation conditions and characterization methodologies (for example: laser wavelength, pulse duration and excitation spot size~\cite{Zimmler2010,Saxena2013OpticallyLasers,Saxena2016}). This is exacerbated by the differences between growth mechanisms (for instance, selective-area growth\cite{Noborisaka2005} versus vapour-liquid-solid\cite{Wagner1964}) and the inclusion of plasmonic\cite{Sidiropoulos2014} or dielectric-enhanced\cite{Sergent2020} emission. 

Tight control of the device fabrication is therefore required to minimise variation in performance and to maximise the useful yield. This is a crucial step towards commercialisation of these miniature technologies, with yield requirements as high as, for example, \SI{99.9999}{\percent} in the microLED industry~\cite{Chen2021s}. Whilst an increasing number of reports now provide statistically robust intra-type results\cite{Wong2023Bottom-upLasers, Church2023HolisticDesign, Church2022OpticalLasers, Jevtics2020CharacterizationSystems}, comparison between nanowire types is rare, causing two significant roadblocks to further nanowire laser development: the lack of a common measurement methodology\cite{Church2022OpticalLasers} to produce comparable data, and the resultant impossibility of obtaining globally meaningful insight from specific studies. 

In this work, we use the principles of experimental data-led discovery~\cite{Merchant2023} to generate an experimental understanding of nanowire laser (NWL) performance, by studying nine different NWL designs with six independent automated experiments, which are detailed in ref\cite{Church2023HolisticDesign}. This includes bright field optical imaging, low power continuous-wave photoluminescence (PL), fluence-dependent PL, emission imaging above the lasing threshold, time-correlated single photon counting (TCSPC) and interferometry, with \StudiedWires NWLs investigated in total. The approach enables the best-in-class NWLs to be identified, a statistically rigorous comparison of the performance of each NWL type and the identification of global trends and important factors that are independent of the NWL design.

\begin{figure*}
    \centering
    \includegraphics[width = \linewidth]{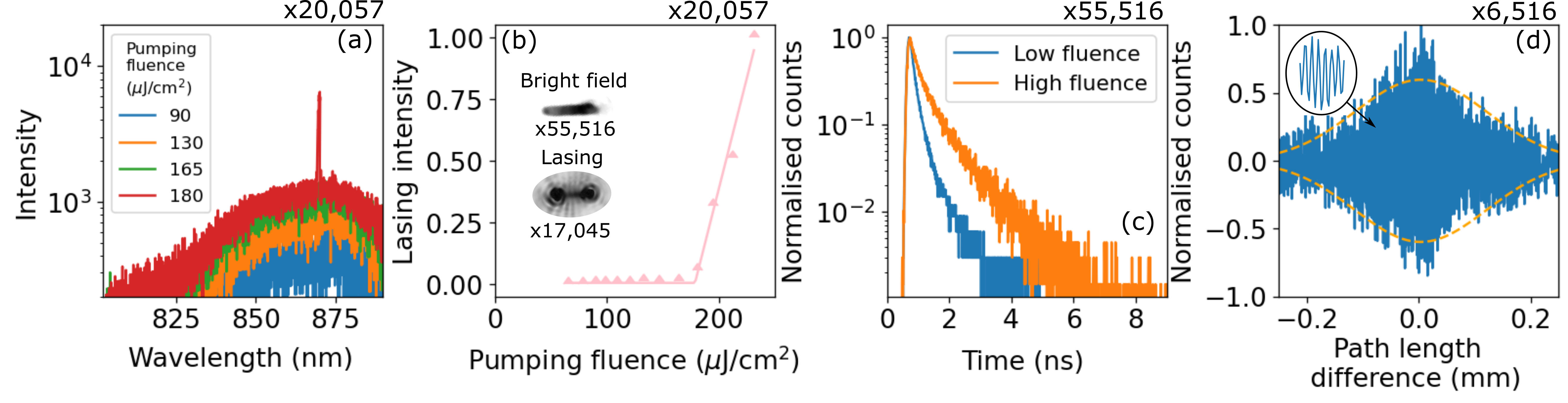}
    \caption{An example of the measurements taken from one individual InP NW, with numbers in the titles indicating the total number of NWs each measurement was performed for. (a) Power dependent photoluminescence spectra showing the emergence of a single lasing mode above a threshold fluence. (b) The light-in-light-out curve for this NWL. The insets show the bright field and lasing emission images. (c) TCSPC PL decays measured below lasing threshold at low fluence and high fluence. (d) Interferogram of the lasing emission, with a Gaussian envelope, used to extract the coherence length. }
    \label{fig:single_wire}
\end{figure*}

A summary of the measurements performed on one individual InP NW are displayed in Figure~\ref{fig:single_wire}. An example low power PL measurement is given in the supporting information. For each measurement, the NW was excited using the output from a PHAROS-ORPHEUS \SI{200}{\femto\second} ultrafast laser-amplifier system with a repetition rate of \SI{100}{\kilo\hertz} and a wavelength of \SI{640}{\nano\meter}. The laser excitation spot was defocussed to uniformly excite the entire NW to achieve maximum gain. Further details of the experimental arrangement can be found in the supporting information and in ref~\cite{Church2023HolisticDesign}. Power-dependent photoluminescence measurements exhibit a broad spontaneous emission peak at all pumping fluences, along with a narrow-band emission peak, above a threshold fluence, that is attributed to stimulated emission. The threshold is identified by analysing the light-in-light-out (LILO) curves of the excitation fluence vs the intensity of the emission, an example of which is shown in Figure~\ref{fig:single_wire}(b). These curves are produced and processed using an established algorithm~\cite{Church2023HolisticDesign}. In short, the emission intensity increases approximately linearly above lasing threshold, and a straight line was fit to this data and extrapolated to $y=0$ in order to estimate a lasing threshold fluence. Using this approach, the threshold for this NW was determined to be \SI{178(3)}{\micro\joule\per\centi\meter\squared}.

Optical imaging measurements for this NW are shown inset in Figure~\ref{fig:single_wire}(b). Bright field images were used to measure the length of the NW, \SI{6}{\micro\meter} in this case. The diameter of the NW was also estimated using these images and found to be \SI{1.3}{\micro\meter}. It is important to note however that this width is similar to the optical resolution of the microscope used (~\SI{0.75}{\micro\meter}), and is therefore likely an over-estimation. This limitation is discussed further in the supporting information.

TCSPC decays measured below the lasing threshold for the example NW are shown in Figure~\ref{fig:single_wire}(c). For this NW, the 1/e decay time was \SI{0.17}{\nano\second} at low fluence, increasing to \SI{0.30}{\nano\second} at a higher fluence. These measured decay times are likely representative of a combination of non-radiative and radiative recombination processes, and as the lifetime increases with increasing fluence, we are unlikely to be in an excitation regime where Auger-Meitner recombination plays an important role for this NW. We therefore suggest that the non-radiative process is related to point defects in the InP and that this fast recombination pathway becomes saturated at higher fluences, thus increasing the effective decay time. This is in agreement with earlier studies, where InP NWs, and SAG grown NWs, have shown low surface recombination (SR) velocities~\cite{Joyce2012}. However, it is important to note that the fluence-dependence varies on a NW-to-NW basis, and thus Auger-Meitner recombination will play some role at high fluences for a subset of the NWs studied. This is discussed further in the supporting information.

Finally, a time-gated interferogram of the lasing emission above threshold was measured, shown in Figure~\ref{fig:single_wire}(d), using an approach detailed in ref~\cite{Church2023HolisticDesign}. A modulation envelope can clearly be observed in the data. The laser coherence length of the radiation was defined as the full width at half maximum (FWHM) of this envelope, which was found to be \SI{0.28}{\milli\meter} for this NW. This coherence length is several orders of magnitude longer than the NW length (\SI{6}{\micro\meter}). This is typical behaviour for the strong waveguides that are formed by NWLs~\cite{Maslov2003_ACS}, and is comparable to previous measurements on alternative materials~\cite{Church2023HolisticDesign}. Using this result, the end-facet reflectivity was calculated to be \SI{59}{\percent} using a previously detailed approach~\cite{Skalsky2020_ACS}. This reflectivity is enhanced beyond the planar Fresnel reflection coefficient for this material (\SI{30}{\percent})~\cite{Adachi1989OpticalInGaAsP}, which may initially be surprising as laser modes extending outside of the small NW cross-section would be anticipated to reduce the reflectivity. However, previous work~\cite{Maslov2003_ACS,Saxena2015} has shown that strong waveguiding effects for certain transverse modes (TE\textsubscript{01} and TM\textsubscript{01}) can result in a reflectivity enhancement comparable to this measurement, for nanowire diameters greater than \SI{300}{\nano\meter}. This is compatible with the diameters of the NWLs studied in this work.

\begin{table*}
    \centering
    \includegraphics[width = 0.95\linewidth]{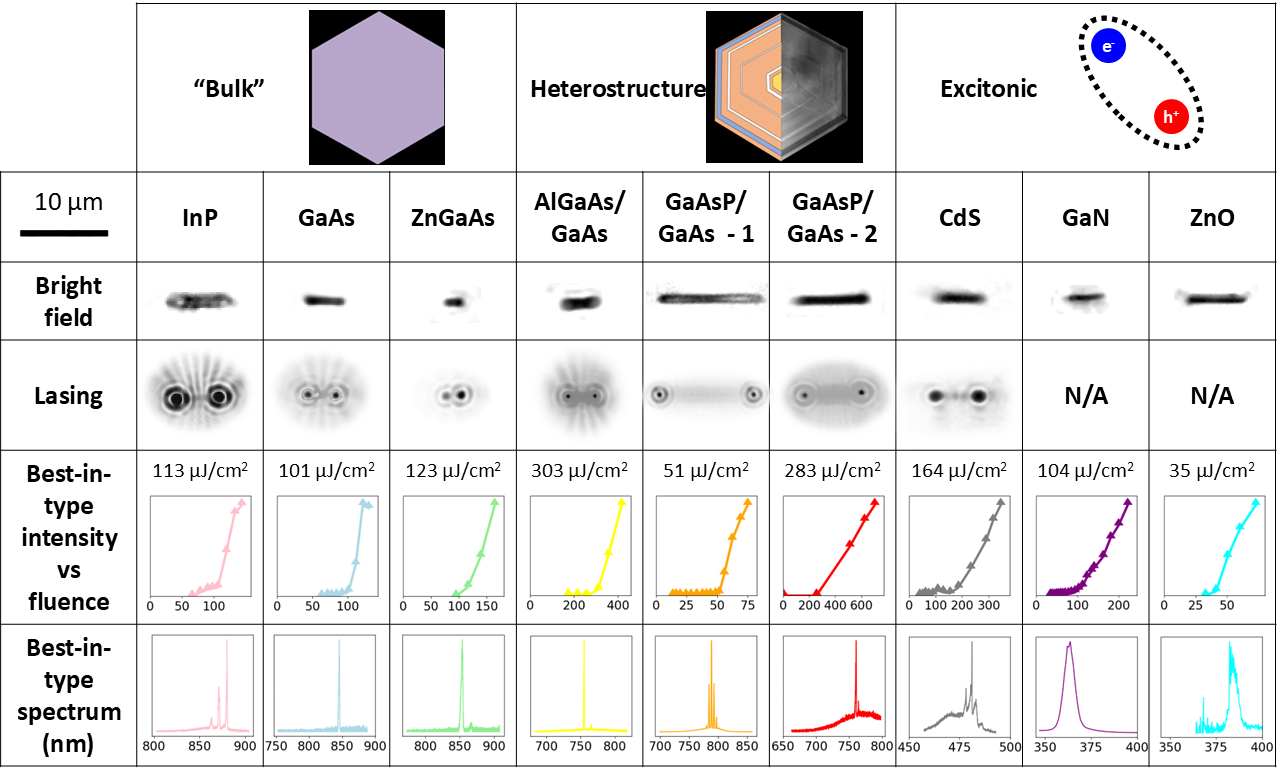}
    \caption{The nine types of nanowires investigated in this study. Bright field optical images are shown for an example NW of each type, along with lasing emission images of the same NWs, demonstrating that the majority of the lasing emission is from the end facets. Emission imaging for GaN and ZnO was not possible due to a lack of sensitivity of the imaging camera to UV light. The lowest threshold NW has been identified for each NW type, and the light-in-light-out curves for each are shown, along with the threshold fluence and the lasing spectrum, measured at the highest fluence in each case. The spectra for the GaN and ZnO nanowires are systematically broadened due to an inferior experimental spectral resolution in the ultraviolet. The heterostructure image has been adapted from ref~\cite{Skalsky2020_ACS}.}
    \label{tab:types}
\end{table*}

A large, multidimensional dataset was generated by performing the measurements in Figure~\ref{fig:single_wire} for NWs of different architectures, which are summarized in Table~\ref{tab:types}. This includes ``bulk" selective-area-grown InP~\cite{Gao2014}, vapour-liquid-solid grown surface-passivated GaAs~\cite{Saxena2013OpticallyLasers}, Zn-doped GaAs~\cite{Alanis2019}, core-shell quantum well (QW) heterostructures AlGaAs/GaAs~\cite{Saxena2016} and GaAsP/GaAs, which has 2 growth batches - 1 and - 2~\cite{Zhang2019_ACS,Skalsky2020_ACS}. These growth batches nominally have the same core/shell structure, and differ only in the duration of the axial NW growth, with batch 1 containing longer NWs on average. Also included are NWs which demonstrate excitonic emission at room temperature due to large exciton binding energies: CdS~\cite{Geburt2012}, GaN~\cite{Jiang2023}, and ZnO~\cite{Borchers2006,Vitale2023}. The NWs were investigated using the same experimental setup as the InP NW in Figure~\ref{fig:single_wire}, and the excitation wavelength was changed depending on the sample. For the ``bulk" and QW samples \SI{640}{\nano\meter} was used, for the CdS NWs, \SI{405}{\nano\meter} was used and for GaN and ZnO, \SI{343}{\nano\meter} was used. A full breakdown of the number of measurements on each NW type is included in the supporting information.  

Table~\ref{tab:types} shows an exemplary bright field optical image for each type of NWL. For comparison, images of the spatial emission profile above lasing threshold for the same wires are also shown in Table~\ref{tab:types}. In this regime, the majority of the light couples out of the end facets of the NW Fabry-Perot cavity. Each end facet acts as an independent, coherent, point source of light and interference fringes between these two sources are observed in the far field~\cite{Saxena2013OpticallyLasers}.

Using this dataset, it is trivial to identify the best-in-type lasers by assessing the lasing thresholds. The LILO curves for these champion NWLs are shown in Table~\ref{tab:types}. The lowest observed threshold was \SI{35(1)}{\micro\joule\per\centi\meter\squared} from a ZnO NWL. This is consistent with measurements using similar excitation conditions on NWs of comparable size and geometry~\cite{Roder2016}. However, this is two orders of magnitude higher than the lowest-reported thresholds in the literature~\cite{Johnson2003}, highlighting the sensitivity of this measurement to the experimental conditions and NW size. 

The GaAsP/GaAs - 1 NWLs also perform favourably, with a lowest threshold of \SI{51(1)}{\micro\joule\per\centi\meter\squared}, which is similar to the best in the literature for this material~\cite{Zhang2019_ACS,Skalsky2020_ACS}. The alternative growth batch GaAsP/GaAs - 2, has a champion threshold which is more than five times higher - emphasising the impact that NW length can have on performance. Despite this, it is interesting to note that the best-in-class thresholds are very similar for all types of "bulk" NWLs studied. This implies that, for the best-performing "bulk" NWLs, the functional performance in the pulsed excitation regime may be limited by the same effects, independent of the material used. This is possibly because the excitation and lasing occurs on timescales faster than many material-dependent carrier recombination processes~\cite{Skalsky2020_ACS}.

The lasing emission spectrum of each best-in-type NW is also shown in Table~\ref{tab:types}, where a variety of behaviours are observed. Emission from multiple longitudinal lasing modes are observed from the InP, \\GaAsP/GaAs~-~1~and CdS example NWs - where the mode spacing is related to the length of the NW cavity [ref].  In contrast, single mode lasing (SML) is observed for the exemplary GaAs, Zn:GaAs, AlGaAs/GaAs and GaAsP/GaAs - 2 NWs at fluences approximately twice that of the threshold fluence. SML is often desired due to the resultant high-spectral purity and is often the result of strong mode selectivity in NWs~\cite{Xiao2011}. his agrees with Table~\ref{tab:types}, where SML is mainly observed in the shorter NW examples.  It should however be noted that SML examples can be found for each NW type.

\begin{figure*}
    \centering
    \includegraphics[width = \linewidth]{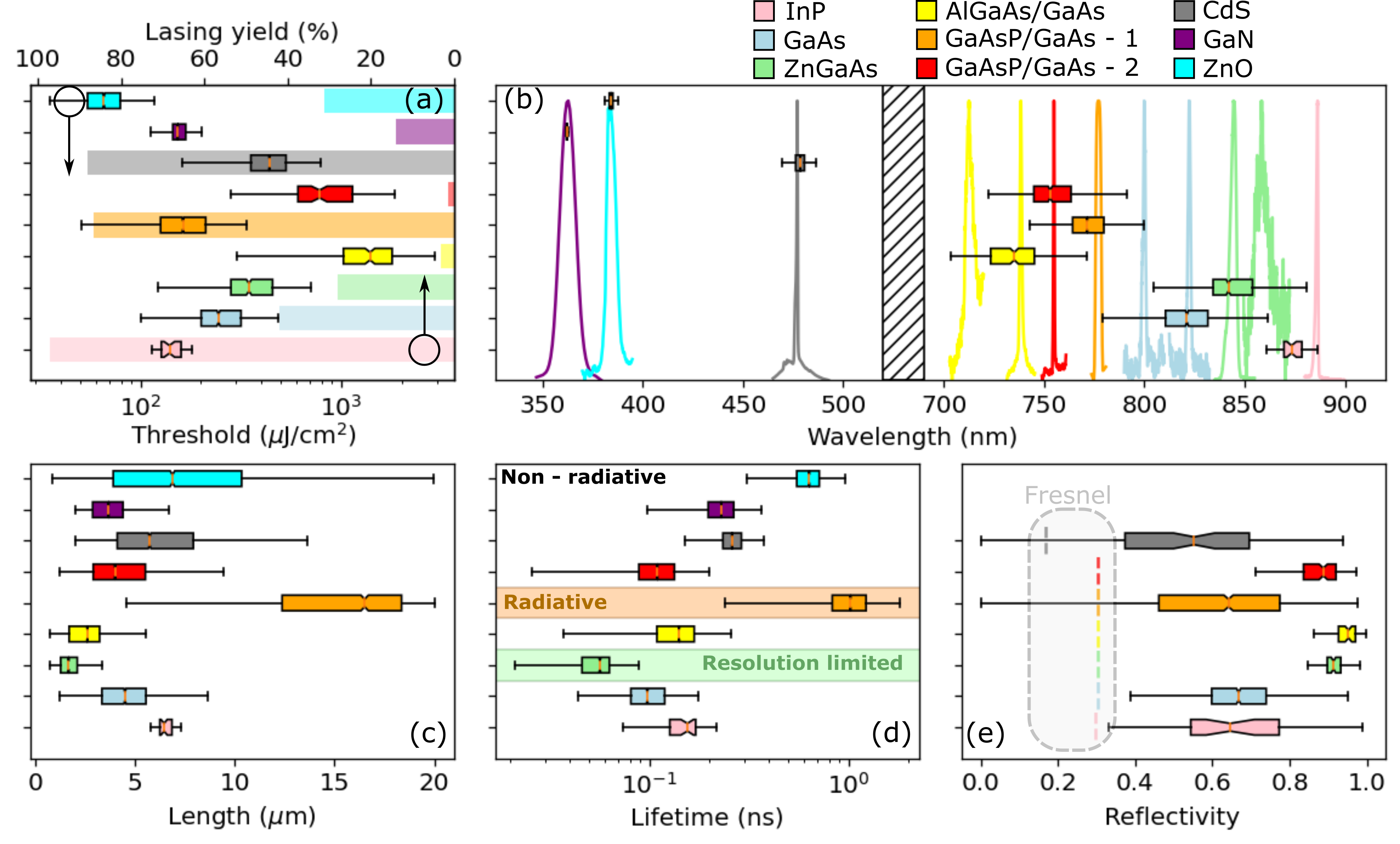}
    \caption{Intra-type statistics. (a) The lasing threshold fluence (box plot) with lasing yield (bar chart) for each NWL type. (b) The peak wavelength of the fundamental lasing mode at threshold. The lasing spectra of selected single-mode NWLs are also shown. (c) Cavity length for all NWs measured from bright field imaging, accurate to \(\approx\)\SI{0.5}{\micro\meter}. (d) The carrier recombination lifetime measured at low fluence, the coloured background indicate dominated radiative (GaAsP/GaAs - 1) and non-radiative decays (white background), and those which are resolution limited (green background). (e) The end-facet reflectivity calculated from the coherence length, the dashed lines are the calculated Fresnel reflection coefficients for a planar surface, assuming the refractive index of InP, GaAs and CdS~\cite{Adachi1989OpticalInGaAsP}. Reflectivity values were not obtained for GaN or ZnO NWLs due to a lack of sensitivity to UV emission for this technique.}
    \label{fig:intratype}
\end{figure*}

The data-led approach enables our study to extend from simple best-in-type measurements (Figure~\ref{fig:single_wire}), through providing robust intra-type measurements of single NWL types (Table~\ref{tab:types}), to understanding global (inter-type) NWL architecture effects. The lasing threshold fluence and the lasing yield (the percentage of NWs with a lasing threshold below their damage threshold) are displayed in Figure~\ref{fig:intratype}(a). The lowest median threshold is seen in ZnO, with \SI{65}{\micro\joule\per\centi\meter\squared}. Furthermore, the GaAsP/GaAs - 1 NWLs, despite having one of the lowest threshold champion devices, demonstrate a large variation in the threshold, such that the median is \SI{163}{\micro\joule\per\centi\meter\squared}, and close to that of the other NWLs. In contrast, the lasing yield for GaAsP/GaAs - 1 NWLs is one of the highest available (\SI{86.5}{\percent}), and is significantly higher than the yield of ZnO NWLs (\SI{31}{\percent}). Clearly the interplay between yield and threshold is not straightforward, and whilst best-in-class devices can offer insights into outlier performance, a statistical approach is required for repeatable inter-class comparison.

The largest median thresholds and threshold variation are seen in the AlGaAs/GaAs and GaAsP/GaAs - 2 core/shell QW NWLs, which have thresholds between 1000 and \SI{2000}{\micro\joule\per\centi\meter\squared}. These NWs also have the lowest yields, just 2 and \SI{3}{\percent}; this is likely because the lasing thresholds are close to the damage threshold of the NWs. Careful NWL heterostructure design is essential to ensure that the QWs overlap well with the transverse lasing modes, and this must be maintained across a population through growth optimisation.

The distribution of the lasing wavelength for each class is shown in Figure~\ref{fig:intratype}(b). These wavelengths depend upon both the laser gain spectrum and the cavity~\cite{Saxena2013OpticallyLasers}. The wavelength variation in each class can be as high as \SI{100}{\nano\meter}: as a result, the full NW population has continuous wavelength coverage across the spectral window between 700 and \SI{880}{\nano\meter}. While this variability presents the most significant obstacle to scale-up, this does provides a means of effectively tuning the wavelength by selecting a NWL with the desired properties\cite{Zapf2019}. To demonstrate this coverage, the lasing spectra for 12 NWLs are shown in Figure~\ref{fig:intratype}(b); these NWLs have been chosen to be single-mode and to change the lasing wavelength by \SI{20}{\nano\meter}, although it is possible to select NWLs based on wavelength with nm-level precision, due to the size of the NW population. This type of data-driven screening is useful as a way to target ideal devices for transfer printing into photonic integrated circuits~\cite{Jevtics2020CharacterizationSystems}.

The distribution of NW lengths are shown in Figure~\ref{fig:intratype}(c). The lengths vary between \SI{0.75}{\micro\meter}, which is the resolution of the optical imaging, and \SI{20}{\micro\meter}. The measured NW diameters were also found to largely correlate with the lengths: the distributions of which are shown in the supporting information. These distributions offer a simple assessment of the uniformity of the growth dynamics and the NWL transfer process~\cite{Alanis2019a}. In this case, the InP NWs stand out as being the most uniform, with a length interquartile range (IQR) of \SI{9}{\percent}; as the only selective-area grown nanowires, this uniformity is a demonstration of the advantage of catalyst-free growth for geometrical homogeneity. The GaAsP/GaAs - 1 length distribution is also distinctive, as this has the largest median length of \SI{16.4}{\micro\meter}, which is more than \SI{10}{\micro\meter} longer than the other heterostructure NW types. This is a strong indication of the importance of the NW length on the performance: longer NWs have a larger volume of gain material and so higher overall gain when excited uniformly~\cite{Alanis2017}. 

Figure 2(d) shows the distributions for the below-threshold photoluminescence lifetime measured from TCSPC decays. The InP, GaAs, AlGaAs/GaAs, GaAsP/GaAs - 2, CdS, GaN and ZnO NWs all exhibit lifetimes that are short when compared with radiative lifetimes of comparable materials~\cite{Gurioli1991,Zad2014,Brandt1998,Niemeyer2019,Keyes1994,Li2001OpticalNm}, with median values between 100 and \SI{640}{\pico\second}, suggesting that non-radiative mechanisms play a role in carrier recombination.  

SR may be anticipated to contribute to this non-radiative recombination, due to the large surface-to-volume ratio of the NWs, however, in many cases the impact of the surface is reduced. For example, InP NWs, grown to the same recipe, have been shown to have SR velocities that are orders of magnitudes below other III/V semiconductors~\cite{Joyce2012}. Furthermore, the GaAs NWs have a shell of AlGaAs which prevents carriers from reaching the surface~\cite{Saxena2013OpticallyLasers}, and a similar effect is achieved for the core/shell AlGaAs/GaAs~\cite{Saxena2016} and GaAsP/GaAs NWs~\cite{Zhang2019_ACS,Skalsky2020_ACS}, as the active QW region is isolated from the surface by the QW barriers. In contrast, the CdS, GaN and ZnO are not surface-passivated. The measurements in this paper do not directly address SR and therefore this cannot be ruled out as a pathway for non-radiative recombination in any of the NW samples.

The Zn-doped GaAs NWs have the fastest median lifetime, \SI{57}{\pico\second}, which is limited by the minimum time resolution of the experiment. Previous studies on the same NWs observed rapid radiative recombination, because of the doping, with a lifetime of 3-\SI{6}{\pico\second}~\cite{Al-Abri2023Sub-PicosecondFramework, Alanis2019}. This is sufficient to out-compete the SR in the NWs. 

Interestingly, the two batches of GaAsP/GaAs NWs have contrasting lifetimes. Batch 1 has the longest median recombination lifetime of \SI{1}{\nano\second}. This batch has been previously studied~\cite{Church2023HolisticDesign}, where it was established that the recombination efficiency is around \SI{39}{\percent}, meaning that the rates of non-radiative and radiative recombination are comparable. The median radiative lifetime is therefore \SI{2.5}{\nano\second}, and the median non-radiative lifetime is \SI{1.6}{\nano\second}. As the impact of SR is reduced in this sample, the non-radiative recombination is likely related to the material quality and the non-radiative defect density. The growth conditions for batch 2 were nominally the same as batch 1 and so it is surprising that the median lifetime is reduced to \SI{0.1}{\nano\second}. This result suggests that the non-radiative rate, and hence the defect density, is higher in the second batch.

The end-facet reflectivity was calculated from measurements of the coherence length\cite{Skalsky2020_ACS} as shown in Figure~\ref{fig:single_wire}(d), and the distributions are shown in Figure~\ref{fig:intratype}(e). This measurement was not possible on the NWs with emission wavelengths less than \SI{400}{\nano\meter} due to limitations in the equipment sensitivity at these wavelengths. The vast majority of NWs have reflectivities that are enhanced above the Fresnel reflection factors for planar surfaces. This is consistent with previous measurements of a single NW class~\cite{Skalsky2020_ACS,Church2023HolisticDesign} and experimentally verifies the theoretical predictions that have underpinned the use of the NW architecture for two decades~\cite{Maslov2003_ACS}. The InP, GaAs, GaAsP/GaAs - 1 and CdS NWs show lower median reflectivities around 0.6 and IQRs up to 0.31. This large degree of variation may be associated with differences in NW diameter or transverse lasing mode - and thus changes in the waveguiding properties~\cite{Maslov2003_ACS}: irregular morphology at the end facets, either due to the NW transfer procedure~\cite{Alanis2019a}, or from an abundance of lattice defects in this region~\cite{Zhang2019a}, may also play a role. The other NWs, Zn:GaAs, AlGaAs/GaAs and GaAsP/GaAs - 2, have high median reflectivities between 0.89 and 0.91 and IQRs up to 0.08. While statistically significant, these findings are likely a result of selection bias in the measurements due to the low yield of these NWLs. Due to the nature of the measurement, reflectivities can only be determined for lasing NWs: thus, high material quality NWLs can operate with poor cavities, while low material quality (poor IQE) NWLs require a high quality cavity to show lasing below their damage threshold.

Whilst analysing each parameter distribution in isolation can provide useful information regarding inhomogeneity in each type of NW, a key benefit of the holistic high-throughput approach is the ability to directly compare the distributions of different parameters for the same NWLs and isolate correlations to investigate physical phenomena. In particular, the parameters that are most important in controlling the lasing threshold can be determined. This is facilitated by the large population sizes in this study: for example, if a correlative analysis is attempted in a small data study, with approximately 100 NWs, an important correlation between two parameters will be obscured by the significant variation in the other parameters, as suggested by the boxplots in Figure~\ref{fig:intratype}.

\begin{figure*}
   \centering
\includegraphics[width = 0.95\linewidth]{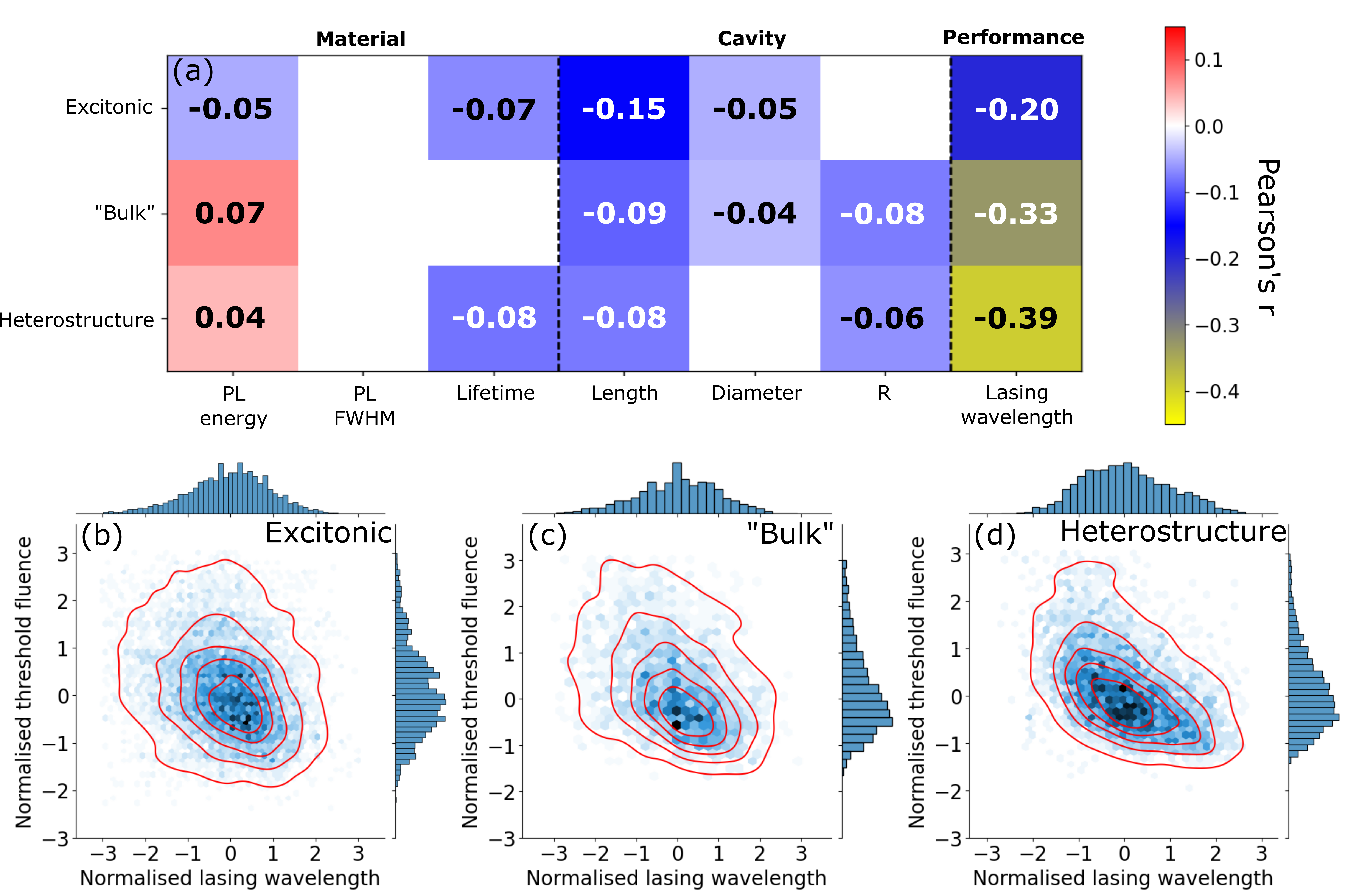}
    \caption{A correlative analysis of the lasing threshold pumping fluence. (a) Statistically significant (with a p-value \(<\)0.05) Pearson's linear correlation coefficients between normalised NW parameters and the lasing pumping threshold, grouped into three NW classes. White indicates parameters where \(p>0.05\) and so there is no statistically significant correlation, whereas red and blue are significant positive and negative correlations respectively. Yellow is representative of a strong negative correlation with \(r<-0.2\). GaAsP/GaAs - 2 and AlGaAs/GaAs NWs are removed from the reflectivity analysis due to selection bias in the distributions. GaN and ZnO NWs are also removed from the reflectivity analysis due to a lack of sensitivity to UV light. Zn:GaAs NWs are removed from the lifetime analysis due to experimental resolution issues. Normalised 2D histograms are also shown, which display the relationship between normalised lasing wavelength and threshold for each class: (b) excitonic, (c) ``bulk" and (d) heterostructure. }
    \label{fig:correlative}
\end{figure*}

To investigate the most important factors that influence the lasing threshold, the datasets were normalised by subtracting the median and dividing by the standard deviation of each distribution for each NW type. The data was then grouped into 3 classes, as defined by Table~\ref{tab:types}: excitonic, ``bulk" and heterostructure. This was necessary to limit the impact of the sparseness, or relatively low population, of some datasets, such as reflectivity (see the supporting information for more information). The linear correlation parameters between the threshold and other key parameters were calculated for each group. This includes material (from PL and lifetime measurements), cavity (dimensions and reflectivity) and performance (lasing wavelength) properties for each group, shown in Figure~\ref{fig:correlative}(a). This provides an at-a-glance means of assessing the performance, averaging over any inter-type variation in NW growths, and enables global trends to be observed that are related to the underlying physics and performance of NWLs. Whilst many of the correlation parameters reported are small (with \(r<0.1\)), this is, at least partially, due to the large degree of variation in multiple other parameters across the NW population. These correlations are statistically significant at the \SI{5}{\percent} level and can therefore be described with confidence. 

These results elucidate the importance of the laser cavity on all classes of NWLs. The NW length has a statistically significant negative correlation with threshold, which, as previously discussed, is due to a increase in the volume of gain material. This dictates a clear route towards optimising the threshold of all types of NWLs: minimise the length variation between each NW. A similar effect is also observed for increasing the diameter of ``bulk" and excitonic NWLs, which is compounded by an additional increase in the optical confinement factor. However, no diameter correlation is observed for heterostructure samples: this suggests that the optical gain is decoupled from the NW diameter and confinement factor, instead depending on the QW-thickness and QW-lasing mode overlap~\cite{Saxena2016}.

The strength of the length correlation reduces from excitonic (r = -0.17), to ``bulk" (r = -0.11) and to heterostructure (r = -0.05) lasers. This suggests that the length has different degree of impact on the threshold for different NW classes. A possible explanation for this is due to differences in distributed losses between the NW classes, which would partially counter-balance the increased gain. For example, for heterostructure NWLs, the overgrowth of the shells can lead to distortions in the NW cross-section shape and consequently enhance the optical losses~\cite{Zhang2019a}.

The cavity reflectivity also plays a crucial role in the performance, which shows a negative correlation with threshold for ``bulk" and heterostructure lasers, indicating that a higher reflectivity reduces the optical losses and, thus, the threshold. No trend is observed for the excitonic NWs, although this may be due to a small sample size (90 wires in total). Another route to reduce variation in NWL performance is therefore to optimise the end facets, either by optimising the NW transfer~\cite{Alanis2019a} or standardising the NW diameter during growth~\cite{Maslov2003_ACS}.

The material properties of the NWLs do not show consistent correlations with threshold. The PL FWHM, an often-used measure of the inhomogeneity in the material~\cite{Alanis2017}, has no statistically significant impact on the threshold. The carrier lifetime also shows no correlation for ``bulk" NWLs, which suggests that, despite the variation in lifetime shown in Fig.~\ref{fig:intratype}(d), there is little significant change in the quantum efficiency of these classes across their populations. In contrast, the heterostructure and excitonic lasers show a negative correlation between lifetime and threshold. This correlation has been discussed in detail previously~\cite{Church2023HolisticDesign}, and is attributed to a variation in the non-radiative recombination rate (and quantum efficiency) which is related to a change in defect density within the gain medium. For these samples, there remains scope to improve the lasing performance by improving the IQE of the active layer by optimising the growth conditions to reduce the defect density.

The PL peak energy in ``bulk" and heterostructure NWs show a positive correlation with threshold. The mechanism for the energy variation is related to composition, strain or heterostructure fluctuations, depending on the sample, and this also leads to shifts in the laser gain spectrum. These correlations are a consequence of the strong negative correlations between lasing wavelength and threshold which have been previously attributed to reabsorption~\cite{Alanis2019} and band-filling effects~\cite{Church2023HolisticDesign}. Interestingly, the excitonic devices show an opposite negative correlation between PL peak energy and threshold. This is consistent with previous studies that demonstrate the importance of strong exciton binding to achieve lasing~\cite{Kilngshirn2007}: NWs with more strongly bound excitons are more likely to have a lower lasing threshold. This is despite the lasing action being sustained by an electron-hole plasma which forms at high fluences~\cite{Versteegh2012}.

The wavelength correlation for each NW class shows the same effects to a different degree. The excitonic devices, shown in Figure~\ref{fig:correlative}(b), have minimal reabsorption since the peak of the gain spectrum lies in the material bandgap, the 3D density of states and excitonic recombination also reduces the impact of band-filling and so the correlation between wavelength and threshold is the lowest of the three classes. The ``bulk" class (Fig.~\ref{fig:correlative}(c)) has more reabsorption, and so the correlation is increased. Finally, the heterostructure NWs (Fig.~\ref{fig:correlative}(d)) exhibit reabsorption into core and cap layers, and will exhibit more band-filling due to a reduced 2D density of states~\cite{Church2023HolisticDesign} and thus have the highest observed correlation.

In summary, we have demonstrated a data-driven experimental approach to study large numbers of micron-scale opto-electronic devices. We have applied this to investigate the properties of \StudiedWires individual NWLs with nine different designs and have extracted parameters that describe how the gain medium, lasing cavity and performance vary across the population. We have shown how the approach can be used to effectively screen for the best-in-class devices and we have quantified the statistical distribution of each parameter and used this to compare the properties of different NWL types in a statistically robust manner. Through a correlative analysis, it was determined that the properties of the laser cavity are consistently a driving factor in the performance of NWLs and this elucidates a route towards achieving homogeneous performance across a population of any type of NWLs: by optimising the NW length and end-facet reflectivity.

\medskip
\textbf{Supporting Information} \par 
Supporting information is available for this paper, which includes the growth details for each type of nanowire, some further experimental details, low power PL spectra and statistics of the NW diameters and lifetimes. This can be found at DOI: ????.

Research data supporting this publication is available at DOI: 10.48420/26403496, and the code to perform the analysis will be made available at DOI: (DOI to be provide in due course), and on github: (URL to be provide in due course).

\medskip
\textbf{Acknowledgements} \par 

This work was funded by UKRI under grants MR/T021519/1, EP/V036343/1 and EP/W002302/1, by the Deutsche Forschungsgemeinschaft within the frame of the collaborative research center CRC 1375 (project C5), by the Australian Research Council, the Australian National Fabrication Facility ACT node, as well as China special Grant (134000-E62201ZJ); Zhejiang Provincial Natural Science Foundation of China (Z24F040009), Zhejiang University Education Foundation Qizhen Scholar Foundation (K20240015), and CMOS Special Program of Zhejiang University (04010000-K2A033208).

CRediT author statement: \textbf{Stephen Church}: Data curation, Formal analysis, Investigation, Methodology, Software, Visualisation, Writing - original draft. \textbf{Francesco Vitale}, \textbf{Aswani Gopakumar}, \textbf{Nikita Gagrani},\textbf{Yunyan Zhang}, \textbf{Nian Jiang}, \textbf{Hark Hoe Tan}, \textbf{Chennupati Jagadish}, \textbf{Huiyun
Liu}, \textbf{Hannah Joyce} and \textbf{Carsten Ronning}: Resources, Writing - review and editing. \textbf{Patrick Parkinson}: Conceptualization, Data curation, Funding acquisition, Methodology, Software, Supervision, Writing - review and editing.

\medskip


\bibliographystyle{MSP}

\renewcommand{\thefigure}{S\arabic{figure}}
\renewcommand{\thetable}{S\arabic{table}}
\renewcommand{\theequation}{S\arabic{equation}}
\setcounter{figure}{0} 
\setcounter{equation}{0} 
\setcounter{table}{0} 

\newpage
\section{Supporting Information}

The nanowires (NWs) studied in this work were grown using various approaches that are summarized in this document. For high-throughput investigation, each sample of NWs was suspended in IPA using ultrasonication for 5s, and transferred to quartz substrates using drop-casting~\cite{Alanis2019}. These NWs were then studied using a number of high-throughput experiments. 

The details of these experiments can be found in a previous publication~\cite{Church2022HolisticNanowires}. In the lasing study, the excitation wavelength was changed between samples to facilitate investigating NWLs with bandgaps between the UV and infra-red. Most samples used the \SI{640}{\nano\meter} output from a PHAROS-ORPHEUS ultrafast laser-amplifier system for excitation. Exceptions to this include the CdS NWLs, which used the laser-amplifier system tuned to \SI{405}{\nano\meter}, and the GaN and ZnO, which used the frequency-tripled output of the PHAROS laser at \SI{343}{\nano\meter}. 

The spot size of the excitation laser depended upon the wavelength used for excitation and varied between 17 and \SI{73}{\micro\meter}. In all cases the excitation spot was larger than the NW, ensuring uniform excitation and maximum gain. The measured lasing thresholds were normalized to the excitation area and reported as a fluence (\si{\micro\joule\per\centi\meter\squared}) and therefore the results are independent of the precise excitation spot size. 

\section{Nanowire growth details}

\subsection{InP}
InP NWs were grown using selective area epitaxy (SAE) with metal-organic chemical vapour deposition (MOCVD). A 150 nm thick SiO\textsubscript{2} mask was deposited on InP (111) wafers, followed by electron beam lithography and dry etching. This produced a mask with dimensions of 100 × \SI{100}{\micro\meter}, which were placed in a close coupled showerhead MOCVD reactor, where trimethyl (TM) Indium and phosphine (PH\textsubscript{3}) were used as precursors with molar flow rates of 6 × 10\textsuperscript{-6} and 4.91 × 10\textsuperscript{-4} \si{\mol\per\minute}, respectively. NWs were grown at \SI{730}{\degreeCelsius} and base pressure of 100 mbar with H\textsubscript{2} as the carrier gas at a total flow rate of \SI{14.5}{\L\per\minute}. More details can be found in reference~\cite{Gao2014}.

\subsection{GaAs}

GaAs NWs were grown on (111) GaAs substrates using a vapor-liquid-solid (VLS) approach via MOCVD in a similar manner to those in ref \cite{Saxena2013OpticallyLasers}. Au particles were used to catalyse the growth, which was achieved using TMGa, TMAl and arsine (AsH\textsubscript{3}) precursors for Ga, Al and As respectively. 

The growth process involved a two-temperature procedure for the GaAs core. Initially, a 1-
minute nucleation step growth at \SI{450}{\degreeCelsius}, followed by 45 minutes of growth at \SI{375}{\degreeCelsius}. Subsequently, the temperature was raised to \SI{750}{\degreeCelsius} in an AsH\textsubscript{3} atmosphere, and an AlGaAs
shell was grown for 0.5 minutes with a TMAl/(TMGa + TMAl) ratio of 0.5. Finally, a GaAs
cap layer was deposited for 1.5 minutes at \SI{750}{\degreeCelsius} to shield the AlGaAs shell from oxidation. More details can be found in reference~\cite{Saxena2013OpticallyLasers}.

\subsection{ZnGaAs}

Zn doped GaAs NWs were grown using a similar Au-catalysed VLS growth approach to the GaAs wires, on (111) GaAs substrates. In this case, the doping was achieved using diethylzinc (DEZn) as a precursor. A capping layer was not included: more information on these wires can be found in reference~\cite{Alanis2019}.

\subsection{AlGaAs/GaAs}

AlGaAs/GaAs quantum well NWs were produced in a similar manner to the GaAs wires: this included a GaAs core which grown using a Au-catalyst mediated two-temperature VLS approach, and an AlGaAs shell layer. This was followed by 8 GaAs/AlGaAs quantum well and barrier layers using an established recipe provided in reference~\cite{Saxena2016}. These wires were also used in a previous optical study~\cite{Alanis2017}.

\subsection{GaAsP/GaAs}

GaAsP/GaAs quantum well NWs were grown using self-catalysed VLS growth in a molecular beam epitaxy (MBE) approach onto p-type (111) Si substrates. A GaAsP NW core was grown was grown at a temperature of \SI{640}{\degreeCelsius} for 90 minutes. The remaining Ga droplet was then consumed by saturating the growth with As and P flux. 3 GaAs/GaAsP quantum well/barrier shell layers were grown at \SI{550}{\degreeCelsius}, followed by outer shell layers of AlGaAsP and GaAsP. Full details of the growth can be found in reference~\cite{Zhang2019ab}.

\subsection{CdS}
CdS NWs were grown using the VLS technique via CVD inside a horizontal-tube vacuum furnace. CdS powder was sublimated at temperatures of \SI{800}{\degreeCelsius} and the vapor was transported by Ar carrier gas towards the growth substrates, which were kept at lower temperatures. The growth substrates (001-Si) were coated with a 10-15 nm Au layer, acting as a catalyst for the growth of CdS NWs, which was carried out for 30 minutes at temperatures between 720 and \SI{600}{\degreeCelsius}. Further details on the growth of these CdS NWs can be found in reference~\cite{Geburt2012}.

\subsection{GaN}

GaN NWs were grown using a self-catalysed approach via MOCVD on c-plane sapphire substrates. The substrates were annealed in H\textsubscript{2} at a temperature of \SI{1060}{\degreeCelsius} and then nitrided at 1080 using NH\textsubscript{3} gas. TMGa and NH\textsubscript{3} were used as precursor gases for a GaN nucleation layer, and NW growth was performed at a temperature of \SI{750}{\degreeCelsius}, using silane (SiH\textsubscript{4}) as an additional precursor to prevent lateral growth. Full growth details can be found in reference \cite{NianJiang2023ComplicationsGrowth}.

\subsection{ZnO}

ZnO NWs were grown via VLS in a similar manner to the CdS samples. In this case, a mixture of ZnO and graphite powder with molar ratio 1:1 was used as the precursor and was sublimated at a temperature of \SI{1050}{\degreeCelsius}. The vapor was transported using Ar + O\textsubscript{2} as the carrier gas and the growth occured at temperatures between 960 and \SI{1150}{\degreeCelsius}. Further details on the growth of ZnO NWs are given in reference~\cite{Borchers2006}.

\section{Low power photoluminescence}

\begin{figure*}[ht]
    \centering
    \includegraphics[width = \linewidth]{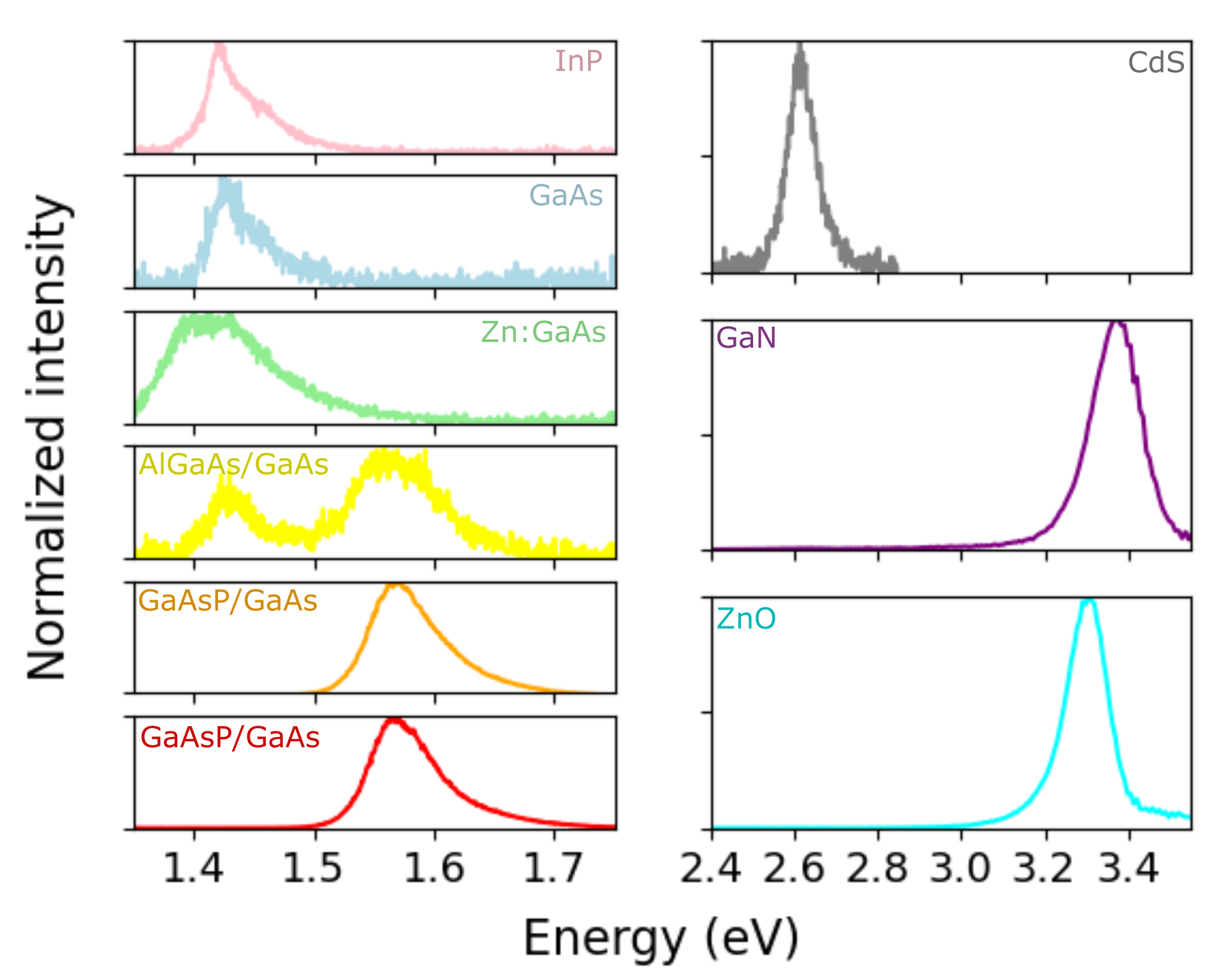}
    \caption{Low power PL spectra for a single NW of each type. }
    \label{fig:PL}
\end{figure*}

Low power photoluminescence (PL) spectra were also measured for all of the \StudiedWires NWs studied in this report. An example spectrum from a single wire of each type is shown in Figure~\ref{fig:PL}. For bulk and heterostructure samples, excitation was achieved using a continuous-wave (CW) HeNe laser with a wavelength of \SI{633}{\nano\meter}, focused to a spot diameter of \SI{4}{\micro\meter} to achieve a power density of \SI{1.4}{\kilo\watt\per\centi\meter\squared}. The CdS NWs were excited using a CW \SI{405}{\nano\meter} laser diode with a spot diameter of \SI{3}{\micro\meter} and a power density of \SI{0.1}{\kilo\watt\per\centi\meter\squared}. The GaN and ZnO were measured using the pulsed \SI{343}{\nano\meter} PHAROS laser, with a spot diameter of \SI{4}{\micro\meter} to achieve a fluence of \SI{100}{\micro\joule\per\centi\meter\squared}.

The PL spectra are compatible with previous studies. InP and GaAs NWs have peak energies of 1.42 and \SI{1.43}{\electronvolt}, and full-widths at half-maximums (FWHMs) of 54 and \SI{64}{\milli\electronvolt} respectively. These results are consistent with band-edge recombination at their nominal bandgaps 1.34~\cite{Gao2014} and \SI{1.42}{\electronvolt}~\cite{Saxena2013OpticallyLasers}. A feature at \SI{1.45}{\electronvolt} is common to both spectra, and is due to the spectral response of the detection system. The Zn-doped GaAs NW shows a PL peak with a similar high-energy side as the GaAs sample. The peak is broadened on the low-energy side, such that the FWHM is increased to \SI{110}{\milli\electronvolt} due to banding of the Zn-related shallow acceptor level which reduces the effective bandgap~\cite{Alanis2019}.

The AlGaAs/GaAs heterostructure NW spectrum has two peaks. The first at \SI{1.43}{\electronvolt} is similar to the GaAs NWs and is due to carrier recombination in the GaAs NW core~\cite{Saxena2016}. The peak at \SI{1.57}{\electronvolt} is due to recombination in the quantum well (QW) and is at a higher energy due to quantum confinement of carriers within the well. The FWHM is relatively large, with a value of \SI{83}{\milli\electronvolt} due to inhomogeneities in the QW. The GaAsP/GaAs NWs both demonstrate emission from the QW layer at a similar energy - but no emission from GaAs as, for these wires, the NW core is composed of GaAsP, which is optically inactive in this measurement~\cite{Zhang2019_ACS,Skalsky2020_ACS}.

The CdS NW shows an emission peak at \SI{2.61}{\electronvolt}, with a FWHM \SI{85}{\milli\electronvolt}. This peak is blue-shifted \SI{200}{\milli\electronvolt} relative to other CdS NWs in the literature~\cite{Geburt2012}. The reason for this is currently unknown, but may be related to bandfilling due to different excitation conditions, or different strain levels or defect states in the material. 

The GaN and ZnO NWs have emission peaks at 3.38 and \SI{3.31}{\electronvolt}, and full-widths at half-maximums (FWHMs) of 143 and \SI{117}{\milli\electronvolt} respectively. These result are comparable with previous observations of near-band-edge emission from these NWs~\cite{Jiang2023,Vitale2023}. For these wires, the recombination occurs below the bandgap due to excitonic effects.

The PL spectra for the NWs were fit using a modified Lasher-Stern-Wurfel (LSW) model for band-edge recombination, which is detailed in a previous publication~\cite{Church2023HolisticDesign}. This enables the PL transition energy and homogeneous broadening term (sigma) to be determined for each NW. These results were included with the full dataset, as well as the FWHM and peak integrated intensity. The excitonic wires were fit with a skewed Gaussian peak, since the LSW model is not valid in this case.

\section{Nanowire population numbers}

\begin{table*}
    \centering
    \includegraphics[width = 0.95\linewidth]{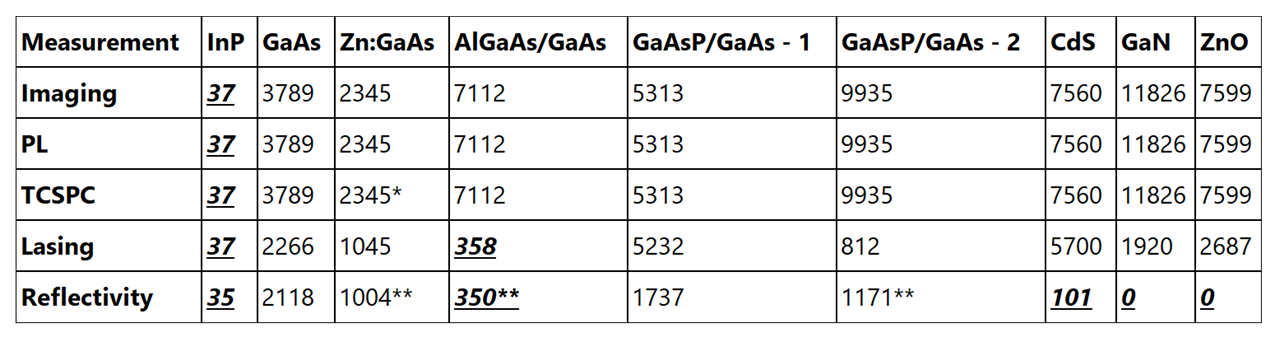}
    \caption{The number of NWs of each type, along with the number of measurements performed on each type. Bold and underlined fields have relatively small populations. Fields with * are resolution limited and those with ** show strong selection bias.}
    \label{tab:types2}
\end{table*}

Table~\ref{tab:types2} shows the number of NWs of each type that were investigated using each experimental technique. When considering this data, challenges exist with the data for some nanowire-types, which have been highlighted in the table.

There are insufficient statistics for some fields to extract correlations. For example, the number of InP NWs studied was only 37 as these were grown in small numbers using SAE. To our knowledge, this is still the most significant statistical evidence for SAE nanowire lasers. Furthermore, the lasing and reflectivity properties of only 358 AlGaAs/GaAs nanowires were studied due to their low yield. The reflectivities of only 101 CdS nanowires were measured due to issues with the sensitivity of the interferometer at short wavelengths. These same sensitivity problems prevented any reflectivity measurements of the GaN and ZnO nanowires. 

The carrier recombination lifetime of the Zn:GaAs nanowires is faster than the resolution of our TCSPC measurement, highlighted by * in the table. We therefore cannot extract any correlation from this data 

Reflectivity measurements can be subject to strong selection bias for nanowires with low yields, such as Zn:GaAs, AlGaAs/GaAs and GaAsP/GaAs - 2, highlighted by ** in the table. This occurs because, for these wires, lasing (and thus the reflectivity measurement) can only occur for high reflectivity, as shown in Figure 2(e). 

The result of these factors is that the correlation analysis is not universally informative on a nanowire-type basis. However, by combining these into 3 classes, weighted by the size of each population, sufficient statistics can be achieved for each experiment and class to draw robust conclusions. The dataset is also publicly available for further post-publication reanalysis. 

\section{Additional nanowire population statistics}

\begin{figure*}
    \centering
    \includegraphics[width = \linewidth]{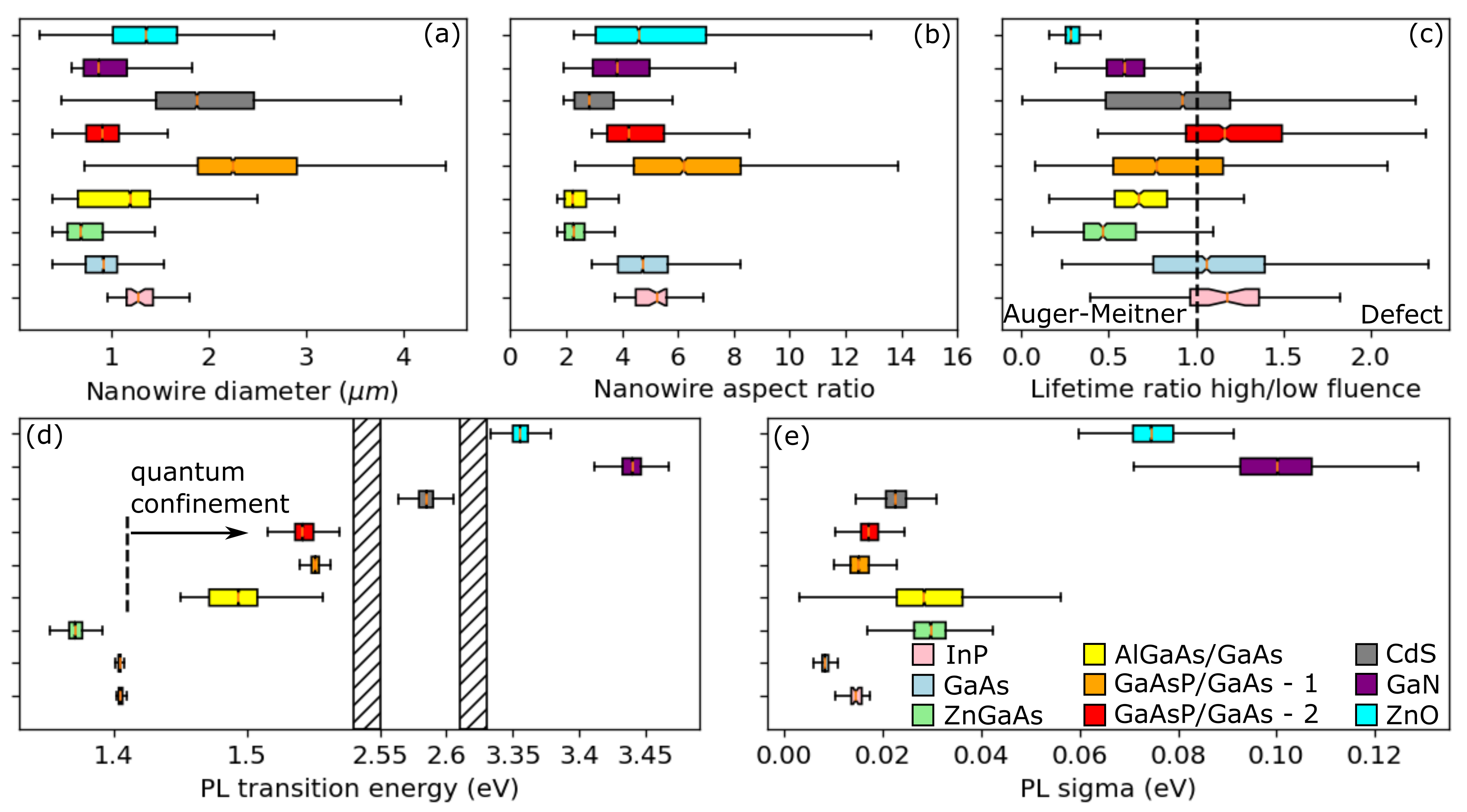}
    \caption{More statistics from the NW dataset. (a) The NW diameter taken from imaging in an optical microscope. (b) The NW aspect ratio, calculated from the length in Figure 2, and the diameter. (c) The ratio of the PL decay times measured at a high fluence close to lasing threshold and a low fluence. The dashed line indicates where the ratio is equal to 1. (d) The PL transition energy for each NW type. The dashed line shows the bandgap of the GaAs NW core for core/shell structures. (e) The PL sigma values for each NW, representative of the inhomogeneity in each NW.}
    \label{fig:additional_stats}
\end{figure*}

Additional statistics for each type of NW are shown in Figure~\ref{fig:additional_stats}. The diameter of each NW was measured using bright-field optical imaging. In most cases, the measured diameter exceeds the diameter measured in previous studies of each growth~\cite{Gao2014,Saxena2013OpticallyLasers,Alanis2019,Saxena2016,Zhang2019ab,Geburt2012,NianJiang2023ComplicationsGrowth,Borchers2006}; this is because the imaging resolution is \SI{0.75}{\micro\meter}. The reported diameters are a convolution of the instrument response function (IRF) and the true diameter. Despite this limitation, Figure~\ref{fig:additional_stats}(a) demonstrates a large spread in the measured diameter, between approximately 0.75 and \SI{4}{\micro\meter}. Since the IRF is unchanged, this variation is due to variation in the true NW diameter - and these results can still be used to assess the diameter variation across the NW population.

The majority of NW types have median measured diameters close to \SI{1}{\micro\meter}, with an inter-quartile range (IQR) of approximately \SI{0.3}{\micro\meter}. Two exceptions to this are the GaAsP/GaAs-1 and CdS samples, which have larger widths and IQRs, of 2.3 and \SI{1.9}{\micro\meter} and 1 and \SI{1}{\micro\meter} respectively. This suggests a higher degree of radial growth for these NWs, along with reduced control of this growth. It is notable that the lowest IQR is for the selective-area epitaxy InP NWs, a result which is comparable with the length distributions in Figure~2(c), and demonstrates the higher degree of growth control provided by this technique over vapour-liquid-solid grown NWs.

Figure~\ref{fig:additional_stats}(b) shows the NW aspect ratio statistics, which is calculated as the ratio of NW length, from Figure 2c, to the NW diameter. In interpreting this data, it is important to note that the length and diameter values are resolution limited. The smaller diameter values are more strongly affected by this, which leads to a systematic bias in the aspect ratio towards smaller values for short and thin wires, and so these aspect ratios represent a lower limit of the true physical aspect ratio. Median values of the aspect ratio vary between 2.5, for the Zn:GaAs and AlGaAs/GaAs samples, and 6.2, for GaAsP/GaAs - 1 sample. The largest observed aspect ratio was 14, and the smallest was 1.8. Qualitatively, the NW types with low thresholds and high yields, such as ZnO, GaAsP/GaAs - 1 and InP tend to have higher aspect ratios, emphasising the impact of the laser cavity on performance. 

The aspect ratio IQR varies between 0.7 and 3.8, the largest values are observed for the ZnO and GaAsP/GaAs - 1 NWs due to their relatively large variation in lengths in Figure~2(c).

Figure~\ref{fig:additional_stats}(c) shows the ratio of the PL decay times measured at a high (close to, and slightly below, lasing threshold) and a low fluence for each NW. In this figure, a ratio of 1 means that the lifetime does not change between the two fluences. A ratio of greater than 1 occurs when the decay time increases at higher fluence, as in the example InP NW in Figure~1. This behaviour is attributed to a saturation of the states associated with Shockley-Read-Hall non-radiative recombination at higher fluences. Alternatively, if the ratio is less than 1, the lifetime reduces at higher fluence, which is attributed to an increase in Auger-Meitner recombination at.

There is a wide spread in fluence-dependent decay time behaviour across the NW population. The InP, GaAs and GaAsP/GaAs-2 types have a median ratio above 1, and the behaviour is therefore mainly dominated by defect-saturation effects, and the impact of Auger-Meitner recombination is limited. In contrast, the Zn:GaAs, AlGaAs/GaAs, GaAsP/GaAs-1, CdS, GaN and ZnO have a median ratio below 1 - suggesting that Auger-Meitner recombination is important at elevated fluences. This effect is largest for the ZnO NWs, which have a median ratio of 0.3. For all other types however, the spread in the data is large, such that there are examples of both types of behaviour in each dataset. 

The variation in this data highlights the complicated nature of fluence dependent decay time behaviour, which is related to a number of competing effects. The photo-excited carrier density will be critical, which will vary on a NW-to-NW basis, depending on the dimensions of the NW, the efficiency of light coupling into the NW and the absorption coefficient and density of states in the active region (QW or bulk). The defect density, Shockley-Read Hall coefficient and Auger-Meitner coefficient in the active region will also determine the overall behaviour. The low fluence decay times are used for the statistical performance comparison in Figure~3, which will be less influenced by Auger-Meitner recombination.

The PL transition energies are shown in Figure~\ref{fig:additional_stats}(c), which mostly follow the lasing wavelength variation in Figure~2, and show similar shifts as the PL spectra in Figure~\ref{fig:PL}. The Zn:GaAs NWs have a smaller bandgap, and a greater bandgap variation, than the bulk samples, due to banding of the acceptor level and variation in the doping density between NWs. The heterostructure QW NWs have a larger bandgap due to quantum confinement effects, and a larger bandgap variation due to QW inhomogeneity. The CdS, GaN and ZnO Nws also have PL bandgaps consistent with the example spectra.

The PL sigma values, a measure of spectral inhomogeneity are shown in Figure~\ref{fig:additional_stats}(d). In a similar result to the PL bandgap variation, median sigma values are comparatively large for the doped and QW samples, which suggests that intra-NW inhomogeneity is large relative to bulk NWs. The sigma values for GaN and ZnO are around \SI{100}{\milli\electronvolt}, and significantly larger than the other materials. This is however due to the fact that a different fit model was used for these materials - a skewed Gaussian, rather than the LSW model.

\medskip
\providecommand{\latin}[1]{#1}
\makeatletter
\providecommand{\doi}
  {\begingroup\let\do\@makeother\dospecials
  \catcode`\{=1 \catcode`\}=2 \doi@aux}
\providecommand{\doi@aux}[1]{\endgroup\texttt{#1}}
\makeatother
\providecommand*\mcitethebibliography{\thebibliography}
\csname @ifundefined\endcsname{endmcitethebibliography}
  {\let\endmcitethebibliography\endthebibliography}{}

\end{document}